\documentclass{article}

\PassOptionsToPackage{numbers,sort,compress}{natbib}
\usepackage[preprint]{neurips_2025}
\usepackage[utf8]{inputenc} 
\usepackage[T1]{fontenc}    
\usepackage{hyperref}       
\usepackage{url}            
\usepackage{booktabs}       
\usepackage{amsfonts}       
\usepackage{nicefrac}       
\usepackage{microtype}      
\usepackage[svgnames]{xcolor}         
\usepackage{graphicx}
\usepackage{amsmath}
\usepackage[normalem]{ulem}

\newif\ifshowcomments
\showcommentstrue

\title{Neural Field Transformations for Hybrid Monte Carlo: Architectural Design and Scaling}

\author{
  Jinchen He \\
  Maryland Center for Fundamental Physics\\
  University of Maryland, College Park\\
  College Park, MD 20742 \\
  \texttt{jinchen@umd.edu} \\
  \And
  Xiao-Yong Jin \\
  Computational Science Division\\
  Argonne National Laboratory\\
  Lemont, IL 60439 \\
  \texttt{xjin@anl.gov} \\
  \AND
  James C. Osborn \\
  Computational Science Division\\
  Argonne National Laboratory\\
  Lemont, IL 60439 \\
  \texttt{osborn@anl.gov} \\
  \And
  Yong Zhao \\
  Physics Division\\
  Argonne National Laboratory\\
  Lemont, IL 60439 \\
  \texttt{yong.zhao@anl.gov} \\
}

\begin{document}

\maketitle

\begin{abstract}
Critical slowing down, where autocorrelation grows rapidly near the continuum limit due to Hybrid Monte Carlo (HMC) moving through configuration space inefficiently, still challenges lattice gauge theory simulations. Combining neural field transformations with HMC (NTHMC) can reshape the energy landscape and accelerate sampling, but the choice of neural architectures has yet to be studied systematically. We evaluate NTHMC on a two-dimensional U(1) gauge theory, analyzing how it scales and transfers to larger volumes and smaller lattice spacing. Controlled comparisons let us isolate architectural contributions to sampling efficiency. Good designs can reduce autocorrelation and boost topological tunneling while maintaining favorable scaling. More broadly, our study highlights emerging design guides, such as wider receptive fields and channel-dependent activations, paving the way for systematic extensions to four-dimensional SU(3) gauge theory.

\end{abstract}

\section{Introduction}

Lattice gauge theory (LGT) simulates gauge fields non-perturbatively by discretizing spacetime on a periodic hypercubic grid. Its flagship application, lattice quantum chromodynamics (QCD) has become the leading non-perturbative tool for studying the strong nuclear interaction.
State-of-the-art calculations use gauge configurations sampled from the Euclidean partition function with Hybrid Monte Carlo (HMC)~\cite{duane1987hybrid}.
Approaching the continuum limit, as the lattice volume grows and lattice spacing shrinks, HMC suffers from \emph{critical slowing down}~\cite{wolff1989critical,alles1996hybrid,del2004critical}---the autocorrelation between gauge configurations in the generated Markov chain increases sharply.
Normalizing flows and related generative models~\cite{albergo2021introduction,abbott2023normalizing} can generate decorrelated samples but struggle with the high dimensionality of realistic lattices~\cite{bulgarelli2025scaling}.
Lüscher’s trivializing map~\cite{luscher2010trivializing}
proposes a construction to transform the physical gauge field to a trivialized field,
performing a change of variables on the path integral.
Practical implementations use invertible smearing steps such as Wilson flow or stout smearing~\cite{yamamoto2025improvement}.
A framework known as neural network transformed HMC (NTHMC)~\cite{jin2022neural,jin2024neural}
constructs gauge-covariant field transformations using machine learning.

This work presents a systematic study of NTHMC, focusing on how neural architectural designs such as residual connections, attention mechanisms, channel-dependent activation, and batch size affect performance and scaling.
Testing on a two-dimensional U(1) gauge theory, we evaluate both small-lattice performance and the transferability to larger volumes and smaller lattice spacing.
Our results suggest that wider receptive fields and channel-dependent activation improve efficiency while maintaining favorable scaling, thereby informing extensions to four-dimensional SU(3) gauge fields and large-scale lattice QCD.

The NTHMC framework uses an invertible, gauge-covariant field transformation $\mathcal{F}: \tilde{U}\mapsto U$ parameterized by a neural network to perform a change of variables in the integration over the Boltzmann weight $\exp(-S(U))$,
leading to an effective action,
\begin{equation}
S_{\mathrm{FT}}(\tilde{U}) = S\big(\mathcal{F}(\tilde{U})\big) - \log\big| \det \mathcal{F}{*}(\tilde{U}) \big| ~,
\end{equation}
where $\mathcal{F}{*}(\tilde{U})$ is the Jacobian matrix of the transformation.
HMC is then performed in the space of auxiliary variables $\tilde{U}$
with the Boltzmann weight $\exp(-S_{\text{FT}}(\tilde{U}))$.

Intuitively, the field transformation acts like a learned preconditioner: it reshapes the energy landscape so that HMC trajectories encounter smoother gradients. We construct the field transformation from gauge-covariant, locally equivariant convolutional neural networks (CNNs). For each link variable $\tilde{U}_{x,\mu}$ (a gauge field element going from site $x$ to $x+\hat{\mu}$), the update takes the form
\begin{equation}
\mathcal{F}: \tilde{U}_{x,\mu} \mapsto U_{x,\mu} = e^{\Pi_{x,\mu}} \tilde{U}_{x,\mu}, \quad
\Pi_{x,\mu} = \sum_{l} \epsilon_{x,\mu,l}\, W_{x,\mu,l} ~.
\end{equation}
The coefficients are given by
$\epsilon_{x,\mu,l} = \phi \!\left[ N_l(X,Y,\dots) \right]$,
with $N_l$ a CNN mapping local gauge-invariant inputs $(X,Y,\dots)$ to the coefficients for the loop functions $W_{x,\mu,l}$, and $\phi$ a bounded activation function (e.g.\ $\tanh$ or $\arctan$) that enforces the invertibility of the transformation. 
$W_{x,\mu,l}$ is the Lie algebra projection of a specific (labeled by $l$) Wilson loop (path ordered products of gauge fields along a closed loop)
that starts and ends at site $x$ and first moves in the $\mu$ direction
(the first on the right of the product).
For the $U(1)$ gauge theory, the projection just takes the imaginary part.

\begin{figure}[ht!]
    \centering
    \includegraphics[width=0.45\linewidth]{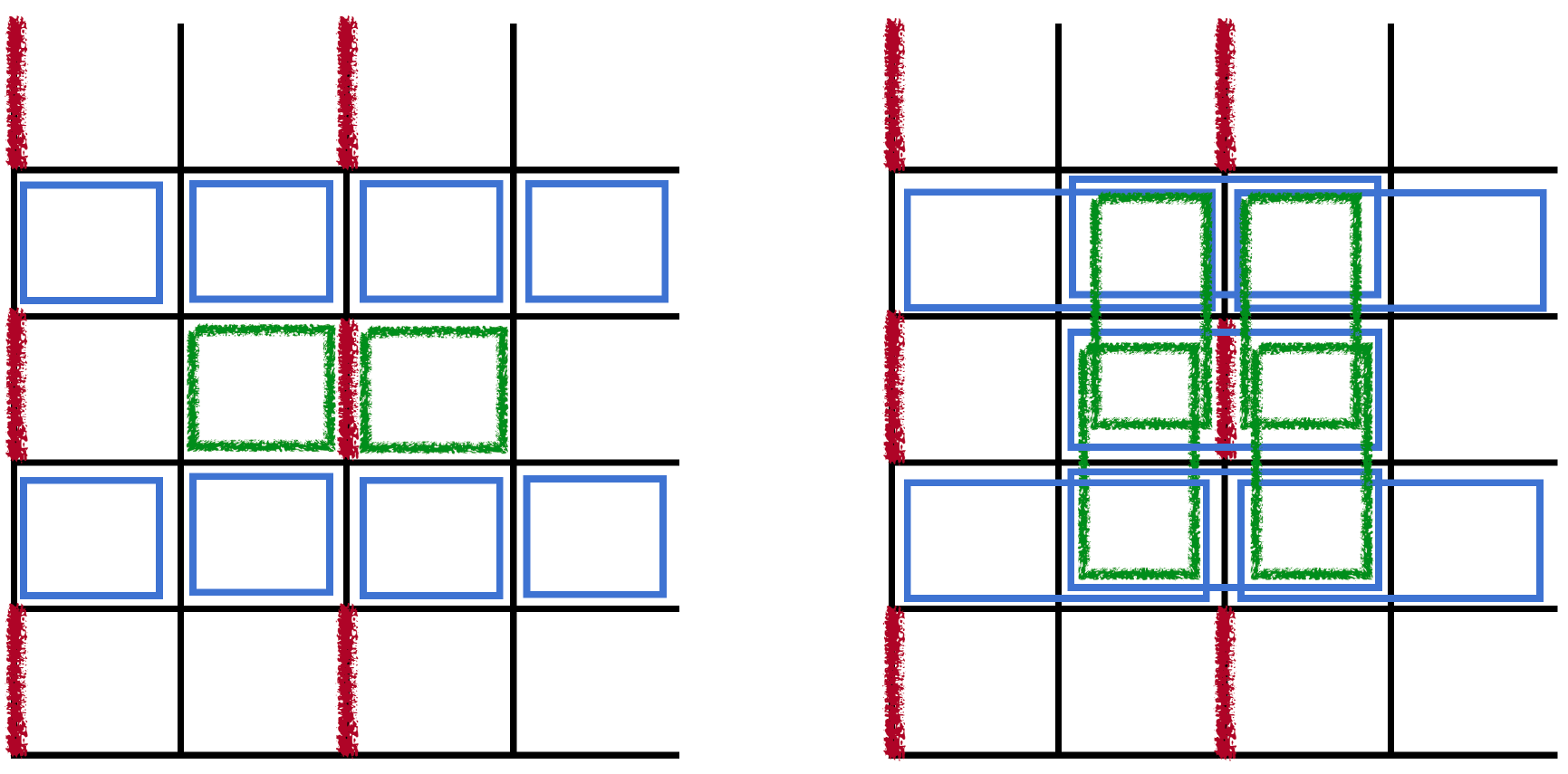}
    \caption{Cropped section of a lattice configuration illustrating the field transformation. Red lines mark one of the eight link subsets updated sequentially, green lines indicate Wilson loops $W_{x,\mu,l}$,
    and blue lines show the gauge-invariant features $(X,Y,\dots)$ used as CNN inputs.}
    \label{fig:subsets}
\end{figure}

As suggested in Ref.~\cite{luscher2010trivializing}, the lattice links are partitioned into eight disjoint subsets and updated sequentially. This ensures that the Jacobian determinant of each update step is reduced to the product of diagonal elements. Moreover, by choosing CNN inputs $(X,Y,\dots)$ to be independent of the links in the active subset, these diagonal elements can be written in a closed form, making the Jacobian determinant explicitly tractable. In this work, we take the $l$ index to run over $1\times1$ plaquettes and $1\times2$ rectangles, as illustrated in Fig.\ref{fig:subsets}, where one link subset (red) is updated, while the corresponding loops (green) and local gauge-invariant CNN inputs (blue) are highlighted.

For U(1) gauge theory,
we use $U=e^{i\theta}$, and the Wilson plaquette action~\cite{wilson1974confinement} as $S(U)$.
The transformation reduces to exponentials of sine terms of plaquette and rectangle loops,
\begin{align}
    \mathcal{F}:~\tilde{U}_{x,\mu} \;\mapsto\; U_{x,\mu} = e^{\sum_{l} \epsilon_{x,\mu,l} \, W_{x,\mu,l}} \, \tilde{U}_{x,\mu} = e^{ \sum_{p} i \epsilon_{x,\mu,p} \sin(\theta_p) + \sum_{r} i \epsilon_{x,\mu,r} \sin(\theta_r)} \tilde{U}_{x,\mu} ~,
\end{align}
with $p$ and $r$ denoting individual plaquette and rectangle loops, respectively, and $\theta_p$, $\theta_r$ the phase angle
of the corresponding Wilson loop.
The Jacobian has the matrix elements
\begin{align}
\mathcal{F}_* (\tilde{U})_{x,\mu,x',\mu'} = 
    \delta_{x,x'} \delta_{\mu,\mu'} \left[
    1 + \sum_p \epsilon_{x,\mu,p} \cos(\theta_p) + \sum_r \epsilon_{x,\mu,r} \cos(\theta_r) \right]~.
\end{align}
In order to have an invertible transformation, the Jacobian determinant may not change sign.
The invertibility condition is then
\begin{align}
    \sum_p |\epsilon_{x,\mu,p}| + \sum_r |\epsilon_{x,\mu,r}| < 1 ~.
    \label{eq:invertibility}
\end{align}

The field transformation is trained on configurations generated from regular HMC by minimizing the transformed action force:
\begin{equation}
    \mathcal{L}(\tilde{U}) = 
    \sum_{p \in \{2,4,6,8\}} 
    \frac{c_{p}}{V^{1/p}} \;
    \left\| \frac{\partial S_{\mathrm{FT}}(\tilde{U})}{\partial \tilde{U}} 
    \right\|_{p} ,
    \label{eq:loss}
\end{equation}
where $V$ is the lattice volume and $c_{p}$ are hyperparameters that weight different force norms, thereby balancing the emphasis between average and peak forces. In our experiments, we set $ c_{2} = c_{4} = c_{6} = c_{8} = 1$. Different from Ref.~\cite{jin2022neural}, which matched the transformed force to that at a larger lattice spacing, we directly minimize the transformed force itself, effectively training the transformation to smooth the action landscape, reduce the stiff directions and improve the sampling efficiency.

\section{Experiments}

Inside the HMC update we employ the 2nd order minimum norm symplectic integrator of Omelyan et al.~\cite{omelyan2003symplectic} with $10$ MD steps per trajectory, tuning the step size to achieve $\sim 80\%$ acceptance. All HMC simulations were verified by checking the plaquette values of the generated configurations, which were found to agree with the theoretical expectations within statistical uncertainties. We benchmark five architectures: a baseline (\texttt{Base}) model similar to the design in Ref.~\cite{jin2022neural}, and four variants (\texttt{Tanh}, \texttt{Resn}, \texttt{Attn}, \texttt{Comb}) introducing different modifications.

\textbf{Base}
The baseline model is a two-layer CNN that processes six input channels (two plaquette, four rectangle). A first $3\times3$ convolution maps them to 12 hidden channels with GELU activation, followed by a second $3\times3$ convolution producing coefficients, which are scaled by $\arctan$ to $(-1/6,1/6)$, satisfying the invertibility condition in Eq.~\ref{eq:invertibility}.
This design yields an effective $5\times5$ receptive field with $\sim$2,000 parameters.

\textbf{Tanh}
Same as \texttt{Base}, but replaces the $\arctan$ scaling with a channel-dependent $\tanh$ activation, ensuring $|\epsilon_{x,\mu,p}| < 0.4$ and $|\epsilon_{x,\mu,r}| < 0.05$, with $\sim 2,000$ parameters.

\textbf{Resn}
Adds two residual blocks (each has two $3\times3$ convolutions with skip connections) before the output, and replaces the second $3\times3$ convolution of \texttt{Base} with a $1\times1$ projection. This enlarges the receptive field to $11\times11$ and increases the parameters to $\sim 6,000$.

\textbf{Attn}
Augments \texttt{Base} with a channel-attention module between the two convolutions, reweighting channels without changing the $5\times5$ receptive field. The parameter count is $\sim 2,100$.

\textbf{Comb}
Combines residual blocks, channel attention, and channel-dependent $\tanh$ activation ($|\epsilon_{x,\mu,p}|<0.4$, $|\epsilon_{x,\mu,r}|<0.05$). The resulting architecture has an $11\times11$ receptive field and $\sim 6,100$ parameters.

All models are trained with the $p$-norm force loss (Eq.\ref{eq:loss}) on $2^{12}$ independent configurations at $\beta=3$, $V=32^2$ using an 80/20 train–test split.
The input gauge coupling, $\beta$, enters in the action $S(U)$, and in this theory, the lattice spacing is proportional to $\beta^{-1/2}$.
The default batch size is $64$, except that \texttt{Base} and \texttt{Comb} are also trained with $32$ (denoted \texttt{Base32}, \texttt{Comb32}) to study batch-size effects. Each model is trained for 32 epochs in PyTorch \cite{paszke2017automatic}, with one single run taking about 12–20 minutes on a single NVIDIA A100 GPU for batch size 64, and roughly twice as long for batch size 32.

To assess the sampling efficiency of regular HMC and NTHMC, we consider two complementary metrics. The first is $\gamma(\delta)$, which measures the relative independence of the configuration~\cite{jin2022neural}:
\begin{align}
    \gamma(\delta) = \frac{1}{1 - \Gamma_t(\delta)}~, \quad \Gamma_t(\delta) \equiv \frac{\left\langle Q_\tau Q_{\tau+\delta}\right\rangle}{\left\langle Q^2\right\rangle} &  =1-\frac{\langle\left(Q_\tau-Q_{\tau+\delta}\right)^2\rangle}{2 V \chi_t^{\infty}(\beta)} ~.
    \label{eq:gamma}
\end{align}
Here $\Gamma_t(\delta)$ is the autocorrelation of the topological charge $Q$ at separation $\delta$ in HMC steps, and $\chi_t^\infty(\beta)$ is the topological susceptibility~\cite{jin2022neural}. The larger $\gamma(\delta)$ indicates stronger autocorrelations and therefore less efficient sampling. 

The second metric is the average topological change $\Delta Q$, the mean absolute shift in topological charge per HMC step, which directly probes the tunneling between sectors; larger values of $\Delta Q$ indicate more frequent topological transitions and thus more efficient sampling. All training and evaluation experiments are repeated with 8 random seeds. Statistical uncertainties are estimated and consistently propagated through derived quantities using Jackknife resampling~\cite{miller1974jackknife} across the seed variability.

\section{Results}

\begin{figure}[ht!]
    \centering
    \includegraphics[width=0.45\linewidth]{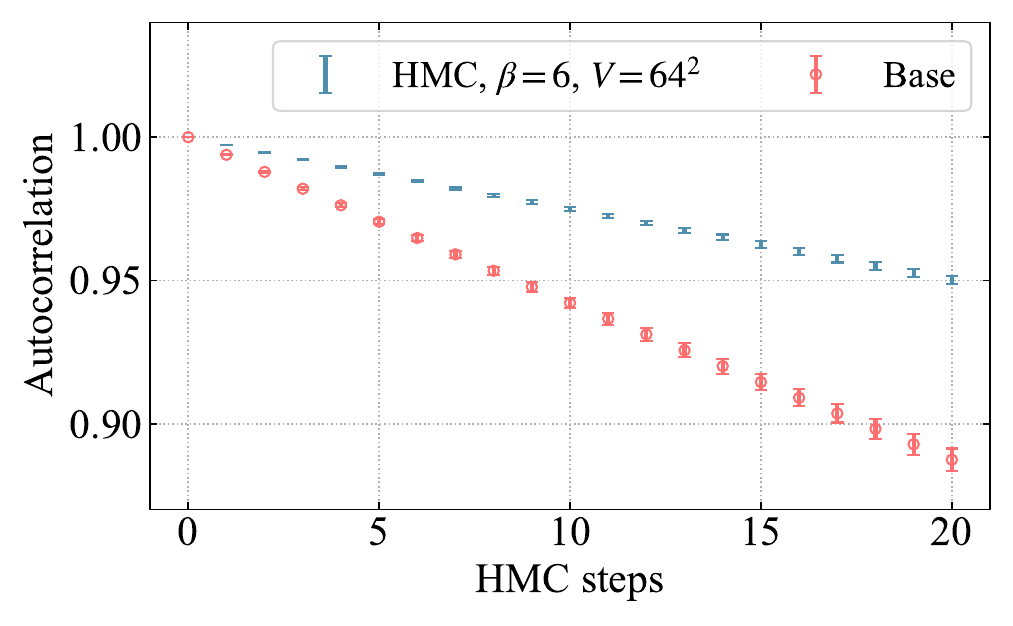}
    \includegraphics[width=0.45\linewidth]{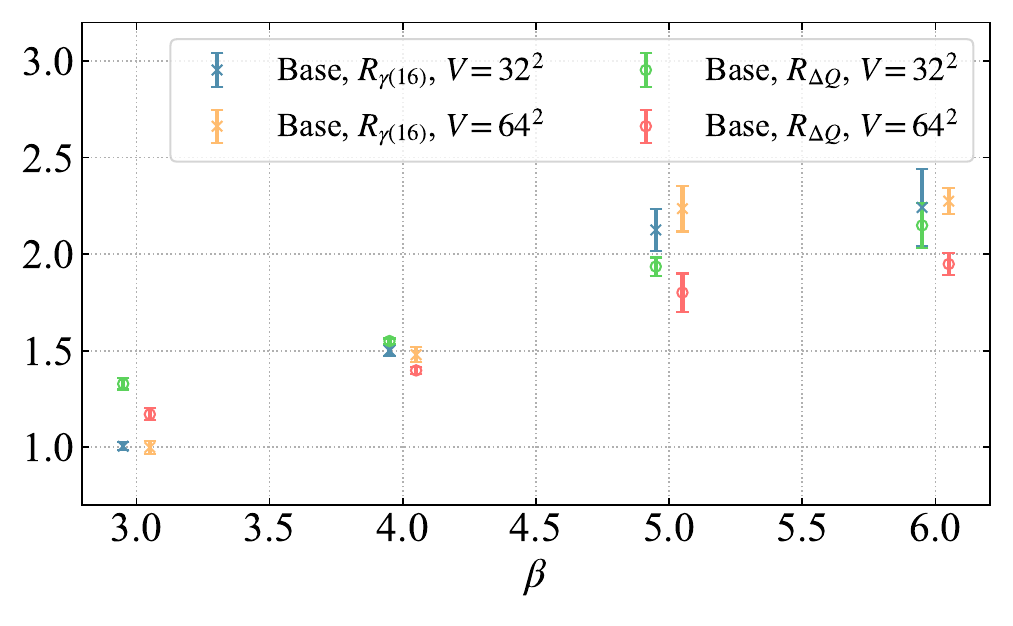}
    \caption{Comparison of regular HMC and NTHMC with the \texttt{Base} model. 
    Left: autocorrelation decay at $\beta=6.0$, $V=64^2$. 
    Right: scaling of two efficiency ratios across $\beta$ and volume, with 
    $R_{\gamma(16)} \equiv \gamma(16)_{\mathrm{HMC}} / \gamma(16)_{\mathrm{Base}}$ 
    and $R_{\Delta Q} \equiv \Delta Q_{\mathrm{Base}} / \Delta Q_{\mathrm{HMC}}$.}
    \label{fig:scaling}
\end{figure}

As shown in the left panel of Fig.~\ref{fig:scaling}, NTHMC with the \texttt{Base} model achieves a markedly faster decay of autocorrelations than regular HMC. The right panel reports the efficiency ratios $R_{\gamma(16)} = \gamma(16)_{\mathrm{HMC}} / \gamma(16)_{\mathrm{Base}}$ and $R_{\Delta Q} = \Delta Q_{\mathrm{Base}} / \Delta Q_{\mathrm{HMC}}$. Despite being trained only at $\beta=3$, $V=32^2$, NTHMC yields even larger relative gains at higher $\beta$ (smaller lattice spacing) and larger volumes, demonstrating favorable scaling and generalization. At $\beta=3$, where regular HMC is already efficient, both ratios remain close to unity.

\begin{figure}[ht!]
    \centering
    \includegraphics[width=0.45\linewidth]{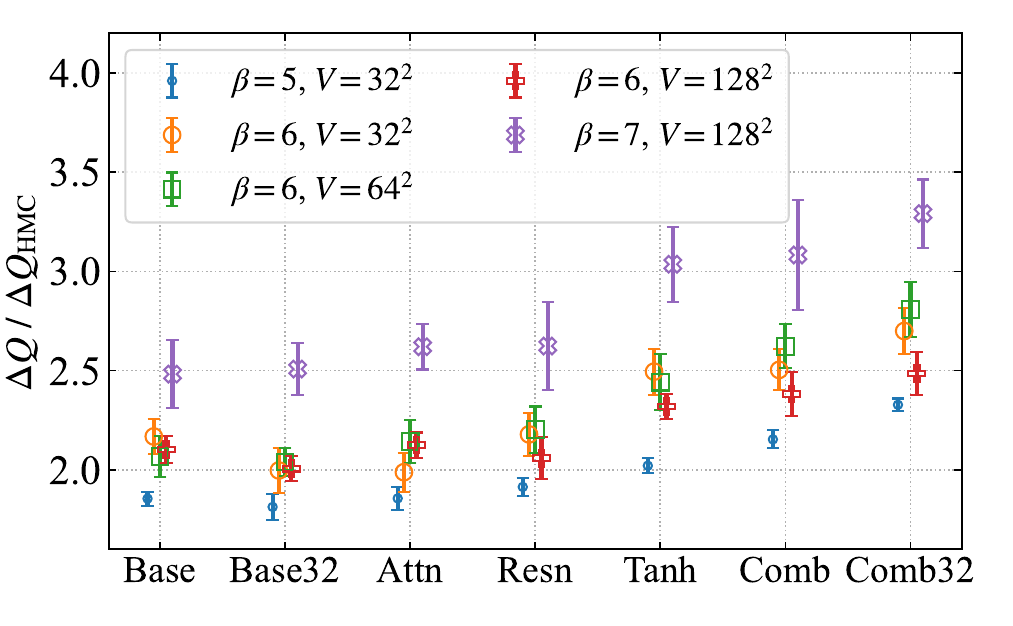}
    \includegraphics[width=0.45\linewidth]{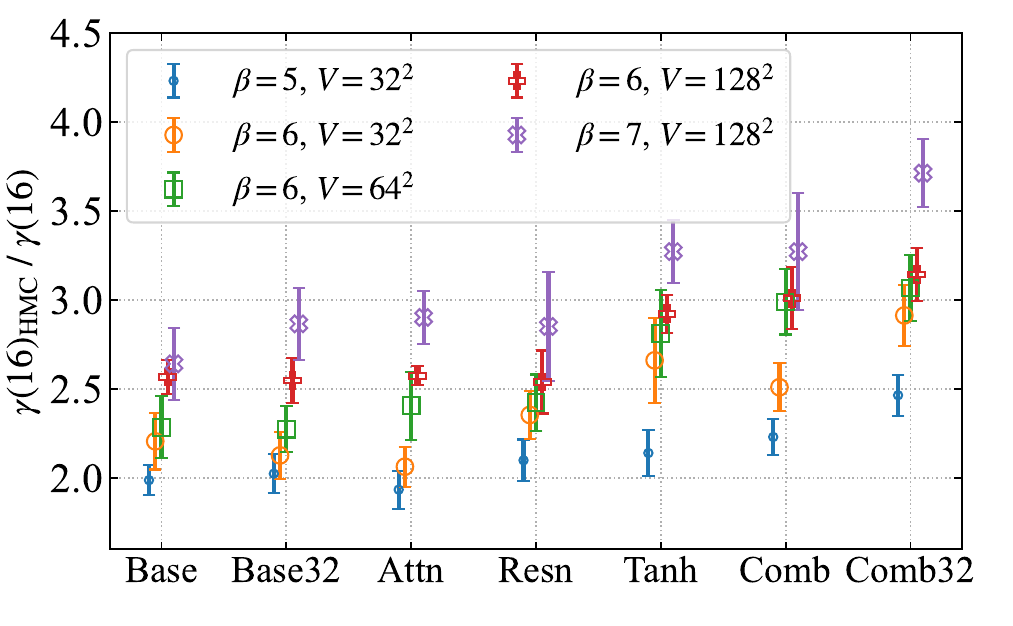}
    \caption{Comparison of NTHMC with various network architectures across different lattice couplings and volumes. The plots show $\Delta Q$ (left panel) and $\gamma(16)$ (right panel), both normalized to standard HMC results.}
    \label{fig:models}
\end{figure}

Fig.~\ref{fig:models} summarizes the performance of different CNN architectures within NTHMC across several lattice couplings and volumes. The comparison includes results at $\beta = 5$, 6, and 7 with lattice sizes $V = 32^2$, $64^2$, and $128^2$, allowing us to examine both coupling and volume dependence relative to standard HMC. Residual blocks and channel-dependent activation lead to improvements over the \texttt{Base} model, and the combined design (\texttt{Comb}) yields the strongest effect, achieving nearly a $40\%$ reduction in $\gamma(16)$ with a corresponding increase in the average topological change $\Delta Q$. Moreover, as the coupling $\beta$ increases, all NTHMC models exhibit a higher sampling efficiency compared to standard HMC, and the efficiency does not show statistically significant changes with increasing lattice volume, indicating that our method can be effectively extended to larger volumes and finer lattices.

Interestingly, the \texttt{Tanh} model achieves competitive gains despite having the same parameter count as \texttt{Base}, which we attribute to the faster saturation of $\tanh$ compared to $\arctan$ and, more importantly, to the larger magnitude of plaquette coefficients $\epsilon_{x,\mu,p}$ enabled by channel-dependent activation. Likewise, residual connections expand the receptive field, which might be expected to hinder transferability, yet we find that an appropriate enlargement does not compromise scaling or generalization and may even be beneficial. A possible explanation is that, as the continuum limit is approached in LGT, infrared (long distance) structures are preserved while ultraviolet (short distance) details become increasingly refined, so the transferable features are precisely the long-range correlations that benefit from a relatively wider receptive field.

Finally, batch size interacts mildly with architecture. Although the baseline shows little sensitivity, the richer \texttt{Comb} model benefits from smaller batches, consistent with gradient noise acting as an implicit regularizer that helps larger capacity models generalize more effectively.

\section{Prospects}

Our preliminary results demonstrate that neural field transformations can substantially reduce autocorrelations, yet several open directions remain. Future work includes exploring integrator choices such as varying the number of MD steps at fixed trajectory length, refining training objectives beyond minimizing the force norms, and systematically analyzing training regimes (e.g. number of epoches, learning rate, training set size and data from different $\beta$ and volumes). Finally, extending from U(1) to SU(3) will be essential to connect this approach with large-scale lattice QCD applications.

\section*{Acknowledgements}

YZ and JH are supported by the U.S. Department of Energy, Office of Science, Office of Nuclear Physics through Contract No.~DE-AC02-06CH11357, within the framework of Scientific Discovery through Advanced Computing (SciDAC) award Fundamental Nuclear Physics at the Exascale and Beyond, and through the American Science Cloud Data Providers Program. YZ and JH also receive partial support by the DOE Office of Science, Office of Nuclear Physics under the umbrella of the Quark-Gluon Tomography (QGT) Topical Collaboration with Award DE-SC0023646. XJ is supported in part by US DOE Contract DE-AC02-06CH11357 and the Scientific Discovery through Advanced Computing (SciDAC) program LAB 22-2580. This research used resources of the Argonne Leadership Computing Facility, a U.S. Department of Energy (DOE) Office of Science user facility at Argonne National Laboratory and is based on research supported by the U.S. DOE Office of Science-Advanced Scientific Computing Research Program, under Contract No. DE-AC02-06CH11357.

\section*{Data Availability}
The code used in this work are available at 
\href{https://github.com/Greyyy-HJC/Scaling_FT_HMC}{https://github.com/Greyyy-HJC/Scaling-FT-HMC}.


\appendix

\bibliographystyle{unsrtnat}
\bibliography{main}

\begin{thebibliography}{15}
\providecommand{\natexlab}[1]{#1}
\providecommand{\url}[1]{\texttt{#1}}
\expandafter\ifx\csname urlstyle\endcsname\relax
  \providecommand{\doi}[1]{doi: #1}\else
  \providecommand{\doi}{doi: \begingroup \urlstyle{rm}\Url}\fi

\bibitem[Duane et~al.(1987)Duane, Kennedy, Pendleton, and Roweth]{duane1987hybrid}
Simon Duane, Anthony~D Kennedy, Brian~J Pendleton, and Duncan Roweth.
\newblock Hybrid monte carlo.
\newblock \emph{Physics letters B}, 195\penalty0 (2):\penalty0 216--222, 1987.

\bibitem[Wolff(1989)]{wolff1989critical}
Ulli Wolff.
\newblock Critical slowing down.
\newblock \emph{Nucl. Phys. B-Proc. Sup}, 17:\penalty0 93--102, 1989.

\bibitem[Alles et~al.(1996)Alles, Boyd, D'Elia, Di~Giacomo, and Vicari]{alles1996hybrid}
B~Alles, G~Boyd, Massimo D'Elia, Adriano Di~Giacomo, and Ettore Vicari.
\newblock Hybrid monte carlo and topological modes of full qcd.
\newblock \emph{Physics Letters B}, 389\penalty0 (1):\penalty0 107--111, 1996.

\bibitem[Del~Debbio et~al.(2004)Del~Debbio, Manca, and Vicari]{del2004critical}
Luigi Del~Debbio, Gian~Mario Manca, and Ettore Vicari.
\newblock Critical slowing down of topological modes.
\newblock \emph{Physics Letters B}, 594\penalty0 (3-4):\penalty0 315--323, 2004.

\bibitem[Albergo et~al.(2021)Albergo, Boyda, Hackett, Kanwar, Cranmer, Racaniere, Rezende, and Shanahan]{albergo2021introduction}
Michael~S Albergo, Denis Boyda, Daniel~C Hackett, Gurtej Kanwar, Kyle Cranmer, S{\'e}bastien Racaniere, Danilo~Jimenez Rezende, and Phiala~E Shanahan.
\newblock Introduction to normalizing flows for lattice field theory.
\newblock \emph{arXiv preprint arXiv:2101.08176}, 2021.

\bibitem[Abbott et~al.(2023)Abbott, Albergo, Botev, Boyda, Cranmer, Hackett, Kanwar, Matthews, Racani{\`e}re, Razavi, et~al.]{abbott2023normalizing}
Ryan Abbott, Michael~S Albergo, Aleksandar Botev, Denis Boyda, Kyle Cranmer, Daniel~C Hackett, Gurtej Kanwar, Alexander~GDG Matthews, S{\'e}bastien Racani{\`e}re, Ali Razavi, et~al.
\newblock Normalizing flows for lattice gauge theory in arbitrary space-time dimension.
\newblock \emph{arXiv preprint arXiv:2305.02402}, 2023.

\bibitem[Bulgarelli et~al.(2025)Bulgarelli, Cellini, and Nada]{bulgarelli2025scaling}
Andrea Bulgarelli, Elia Cellini, and Alessandro Nada.
\newblock Scaling of stochastic normalizing flows in su (3) lattice gauge theory.
\newblock \emph{Physical Review D}, 111\penalty0 (7):\penalty0 074517, 2025.

\bibitem[L{\"u}scher(2010)]{luscher2010trivializing}
Martin L{\"u}scher.
\newblock Trivializing maps, the wilson flow and the hmc algorithm.
\newblock \emph{Communications in mathematical physics}, 293\penalty0 (3):\penalty0 899--919, 2010.

\bibitem[Yamamoto et~al.(2025)Yamamoto, Boyle, Izubuchi, Jin, Lehner, and Matsumoto]{yamamoto2025improvement}
Shuhei Yamamoto, Peter Boyle, Taku Izubuchi, Luchang Jin, Christoph Lehner, and Nobuyuki Matsumoto.
\newblock Improvement in autocorrelation times measured by the master-field technique using field transformation hmc in 2+ 1 domain wall fermion simulations.
\newblock \emph{arXiv preprint arXiv:2502.05452}, 2025.

\bibitem[Jin(2022)]{jin2022neural}
Xiao-yong Jin.
\newblock Neural network field transformation and its application in hmc.
\newblock \emph{arXiv preprint arXiv:2201.01862}, 2022.

\bibitem[Jin(2024)]{jin2024neural}
Xiao-Yong Jin.
\newblock Neural network gauge field transformation for 4d su (3) gauge fields.
\newblock \emph{arXiv preprint arXiv:2405.19692}, 2024.

\bibitem[Wilson(1974)]{wilson1974confinement}
Kenneth~G Wilson.
\newblock Confinement of quarks.
\newblock \emph{Physical review D}, 10\penalty0 (8):\penalty0 2445, 1974.

\bibitem[Omelyan et~al.(2003)Omelyan, Mryglod, and Folk]{omelyan2003symplectic}
IP~Omelyan, IM~Mryglod, and R~Folk.
\newblock Symplectic analytically integrable decomposition algorithms: classification, derivation, and application to molecular dynamics, quantum and celestial mechanics simulations.
\newblock \emph{Computer Physics Communications}, 151\penalty0 (3):\penalty0 272--314, 2003.

\bibitem[Paszke et~al.(2017)Paszke, Gross, Chintala, Chanan, Yang, DeVito, Lin, Desmaison, Antiga, and Lerer]{paszke2017automatic}
Adam Paszke, Sam Gross, Soumith Chintala, Gregory Chanan, Edward Yang, Zachary DeVito, Zeming Lin, Alban Desmaison, Luca Antiga, and Adam Lerer.
\newblock Automatic differentiation in pytorch.
\newblock 2017.

\bibitem[Miller(1974)]{miller1974jackknife}
Rupert~G Miller.
\newblock The jackknife-a review.
\newblock \emph{Biometrika}, 61\penalty0 (1):\penalty0 1--15, 1974.

\end{thebibliography}

\end{document}